\documentclass[conference]{IEEEtran}
\usepackage{amsmath,amsfonts}
\usepackage{algorithmic}
\usepackage{array}
\usepackage{textcomp}
\usepackage{stfloats}
\usepackage{url}
\usepackage{verbatim}
\usepackage{graphicx}
\usepackage{xspace}
\usepackage{amsmath}
\usepackage{svg}
\def\BibTeX{{\rm B\kern-.05em{\sc i\kern-.025em b}\kern-.08em
    T\kern-.1667em\lower.7ex\hbox{E}\kern-.125emX}}
\usepackage{balance}
\usepackage[acronyms,nonumberlist,nopostdot,nomain,nogroupskip,acronymlists={hidden}]{glossaries}
\usepackage{soul}
\usepackage{color}
\newglossary[algh]{hidden}{acrh}{acnh}{Hidden Acronyms}
\newcommand{\framework}{AUGUSTE\xspace}
\usepackage{tikz}
\usetikzlibrary{positioning, arrows.meta}
\newcommand{\cnum}[1]{\tikz[baseline=(c.base)]{\node[draw,circle,inner sep=0.5pt,line width=0.4pt,minimum size=1.6ex,font=\scriptsize](c){#1};}}
\usepackage{pgfplots}
\pgfplotsset{compat=1.18}  
\newlength\fheight
\newlength\fwidth
\newacronym{isac}{ISAC}{Integrated Sensing and Communication}
\newacronym{tpc}{TPC}{Transmit Power Control}
\newacronym{re}{RE}{Resource Element}
\newacronym{ewma}{EWMA}{Exponentially Weighted Moving Average}
\newacronym{gpfb}{GPFB}{Generalized Proportional Fair Buffer}
\newacronym{drb}{DRB}{Data Radio Bearer}
\newacronym{hol}{D-HoL}{Head-of-Line Delay}
\newacronym{bcqi}{BCQI}{Best CQI}
\newacronym{ft}{FT}{Fair Throughput}
\newacronym{rr}{RR}{Round-Robin}
\newacronym{pf}{PF}{Proportional Fair}
\newacronym{x5g}{X5G}{}
\newacronym{ibs}{IBS}{Intent-Based Scheduling}
\newacronym{rapp}{rApp}{RAN Application}
\newacronym{xapp}{xApp}{eXtended Application}
\newacronym{dapp}{dApp}{Distributed Application}
\newacronym{ibn}{IBN}{Intent-Based Networking}
\newacronym{dlsch}{DLSCH}{Downlink Shared Channel}
\newacronym{ocr}{OCR}{Optical Character Recognition}
\newacronym{3gpp}{3GPP}{3rd Generation Partnership Project}
\newacronym{sriov}{SR-IOV}{Single Root I/O Virtualization}
\newacronym{vf}{VF}{Vitual Functions}
\newacronym{4g}{4G}{4th generation}
\newacronym{5g}{5G}{5th generation}
\newacronym{6g}{6G}{6th generation}
\newacronym{5gc}{5GC}{5G Core}
\newacronym{adc}{ADC}{Analog to Digital Converter}
\newacronym{aerpaw}{AERPAW}{Aerial Experimentation and Research Platform for Advanced Wireless}
\newacronym{ai}{AI}{Artificial Intelligence}
\newacronym{aimd}{AIMD}{Additive Increase Multiplicative Decrease}
\newacronym{am}{AM}{Acknowledged Mode}
\newacronym{amc}{AMC}{Adaptive Modulation and Coding}
\newacronym{amf}{AMF}{Access and Mobility Management Function}
\newacronym{aops}{AOPS}{Adaptive Order Prediction Scheduling}
\newacronym{api}{API}{Application Programming Interface}
\newacronym{apn}{APN}{Access Point Name}
\newacronym{ap}{AP}{Application Protocol}
\newacronym{aqm}{AQM}{Active Queue Management}
\newacronym{ausf}{AUSF}{Authentication Server Function}
\newacronym{avc}{AVC}{Advanced Video Coding}
\newacronym{awgn}{AGWN}{Additive White Gaussian Noise}
\newacronym{balia}{BALIA}{Balanced Link Adaptation Algorithm}
\newacronym{bbu}{BBU}{Base Band Unit}
\newacronym{bdp}{BDP}{Bandwidth-Delay Product}
\newacronym{ber}{BER}{Bit Error Rate}
\newacronym{bf}{BF}{Beamforming}
\newacronym{bler}{BLER}{Block Error Rate}
\newacronym{brr}{BRR}{Bayesian Ridge Regressor}
\newacronym{bs}{BS}{Base Station}
\newacronym{bsr}{BSR}{Buffer Status Report}
\newacronym{bss}{BSS}{Business Support System}
\newacronym{ca}{CA}{Carrier Aggregation}
\newacronym{caas}{CaaS}{Connectivity-as-a-Service}
\newacronym{cc}{CC}{Congestion Control}
\newacronym{ccid}{CCID}{Congestion Control ID}
\newacronym{cco}{CC}{Carrier Component}
\newacronym{cd}{CD}{Continuous Delivery}
\newacronym{cdd}{CDD}{Cyclic Delay Diversity}
\newacronym{cdf}{CDF}{Cumulative Distribution Function}
\newacronym{cdn}{CDN}{Content Distribution Network}
\newacronym{cli}{CLI}{Command-line Interface}
\newacronym{cn}{CN}{Core Network}
\newacronym{codel}{CoDel}{Controlled Delay Management}
\newacronym{comac}{COMAC}{Converged Multi-Access and Core}
\newacronym{cord}{CORD}{Central Office Re-architected as a Datacenter}
\newacronym{cornet}{CORNET}{COgnitive Radio NETwork}
\newacronym{cosmos}{COSMOS}{Cloud Enhanced Open Software Defined Mobile Wireless Testbed for City-Scale Deployment}
\newacronym{cots}{COTS}{Commercial Off-the-Shelf}
\newacronym{cp}{CP}{Control Plane}
\newacronym{cyp}{CP}{Cyclic Prefix}
\newacronym{up}{UP}{User Plane}
\newacronym{cpu}{CPU}{Central Processing Unit}
\newacronym{cqi}{CQI}{Channel Quality Information}
\newacronym{cr}{CR}{Cognitive Radio}
\newacronym{cran}{CRAN}{Cloud RAN}
\newacronym{crs}{CRS}{Cell Reference Signal}
\newacronym{csi}{CSI}{Channel State Information}
\newacronym{csirs}{CSI-RS}{Channel State Information - Reference Signal}
\newacronym{cu}{CU}{Central Unit}
\newacronym{cubb}{cuBB}{CUDA Baseband}
\newacronym{d2tcp}{D$^2$TCP}{Deadline-aware Data center TCP}
\newacronym{d3}{D$^3$}{Deadline-Driven Delivery}
\newacronym{dac}{DAC}{Digital to Analog Converter}
\newacronym{dag}{DAG}{Directed Acyclic Graph}
\newacronym{das}{DAS}{Distributed Antenna System}
\newacronym{dash}{DASH}{Dynamic Adaptive Streaming over HTTP}
\newacronym{dc}{DC}{Dual Connectivity}
\newacronym{dccp}{DCCP}{Datagram Congestion Control Protocol}
\newacronym{dce}{DCE}{Direct Code Execution}
\newacronym{dci}{DCI}{Downlink Control Information}
\newacronym{dctcp}{DCTCP}{Data Center TCP}
\newacronym{dl}{DL}{Downlink}
\newacronym{dmr}{DMR}{Deadline Miss Ratio}
\newacronym{dmrs}{DMRS}{DeModulation Reference Signal}
\newacronym{drlcc}{DRL-CC}{Deep Reinforcement Learning Congestion Control}
\newacronym{drs}{DRS}{Discovery Reference Signal}
\newacronym{du}{DU}{Distributed Unit}
\newacronym{e2e}{E2E}{end-to-end}
\newacronym{earfcn}{EARFCN}{E-UTRA Absolute Radio Frequency Channel Number}
\newacronym{ecaas}{ECaaS}{Edge-Cloud-as-a-Service}
\newacronym{ecn}{ECN}{Explicit Congestion Notification}
\newacronym{edf}{EDF}{Earliest Deadline First}
\newacronym{embb}{eMBB}{Enhanced Mobile Broadband}
\newacronym{empower}{EMPOWER}{EMpowering transatlantic PlatfOrms for advanced WirEless Research}
\newacronym{enb}{eNB}{evolved Node Base}
\newacronym{endc}{EN-DC}{E-UTRAN-\gls{nr} \gls{dc}}
\newacronym{epc}{EPC}{Evolved Packet Core}
\newacronym{eps}{EPS}{Evolved Packet System}
\newacronym{es}{ES}{Edge Server}
\newacronym{etsi}{ETSI}{European Telecommunications Standards Institute}
\newacronym[firstplural=Estimated Times of Arrival (ETAs)]{eta}{ETA}{Estimated Time of Arrival}
\newacronym{eutran}{E-UTRAN}{Evolved Universal Terrestrial Access Network}
\newacronym{faas}{FaaS}{Function-as-a-Service}
\newacronym{fapi}{FAPI}{Functional Application Platform Interface}
\newacronym{fdd}{FDD}{Frequency Division Duplexing}
\newacronym{fdm}{FDM}{Frequency Division Multiplexing}
\newacronym{fdma}{FDMA}{Frequency Division Multiple Access}
\newacronym{fed4fire}{FED4FIRE+}{Federation 4 Future Internet Research and Experimentation Plus}
\newacronym{fir}{FIR}{Finite Impulse Response}
\newacronym{fit}{FIT}{Future \acrlong{iot}}
\newacronym{fpga}{FPGA}{Field Programmable Gate Array}
\newacronym{fr2}{FR2}{Frequency Range 2}
\newacronym{fs}{FS}{Fast Switching}
\newacronym{fscc}{FSCC}{Flow Sharing Congestion Control}
\newacronym{ftp}{FTP}{File Transfer Protocol}
\newacronym{fw}{FW}{Flow Window}
\newacronym{ge}{GE}{Gaussian Elimination}
\newacronym{gnb}{gNB}{Next Generation Node Base}
\newacronym{gop}{GOP}{Group of Pictures}
\newacronym{gpr}{GPR}{Gaussian Process Regressor}
\newacronym{gpu}{GPU}{Graphics Processing Unit}
\newacronym{gtp}{GTP}{GPRS Tunneling Protocol}
\newacronym{gtpc}{GTP-C}{GPRS Tunnelling Protocol Control Plane}
\newacronym{gtpu}{GTP-U}{GPRS Tunnelling Protocol User Plane}
\newacronym{gtpv2c}{GTPv2-C}{\gls{gtp} v2 - Control}
\newacronym{gw}{GW}{Gateway}
\newacronym{harq}{HARQ}{Hybrid Automatic Repeat reQuest}
\newacronym{hetnet}{HetNet}{Heterogeneous Network}
\newacronym{hh}{HH}{Hard Handover}
\newacronym{hqf}{HQF}{Highest-quality-first}
\newacronym{hss}{HSS}{Home Subscription Server}
\newacronym{http}{HTTP}{HyperText Transfer Protocol}
\newacronym{ia}{IA}{Initial Access}
\newacronym{iab}{IAB}{Integrated Access and Backhaul}
\newacronym{ic}{IC}{Incident Command}
\newacronym{ietf}{IETF}{Internet Engineering Task Force}
\newacronym{imsi}{IMSI}{International Mobile Subscriber Identity}
\newacronym{imt}{IMT}{International Mobile Telecommunication}
\newacronym{iot}{IoT}{Internet of Things}
\newacronym{ip}{IP}{Internet Protocol}
\newacronym{itu}{ITU}{International Telecommunication Union}
\newacronym{kpi}{KPI}{Key Performance Indicator}
\newacronym{kpm}{KPM}{Key Performance Measurement}
\newacronym{kvm}{KVM}{Kernel-based Virtual Machine}
\newacronym{los}{LOS}{Line-of-Sight}
\newacronym{lsm}{LSM}{Link-to-System Mapping}
\newacronym{lstm}{LSTM}{Long Short Term Memory}
\newacronym{lte}{LTE}{Long Term Evolution}
\newacronym{lxc}{LXC}{Linux Container}
\newacronym{m2m}{M2M}{Machine to Machine}
\newacronym{mac}{MAC}{Medium Access Control}
\newacronym{manet}{MANET}{Mobile Ad Hoc Network}
\newacronym{mano}{MANO}{Management and Orchestration}
\newacronym{mc}{MC}{Multi-Connectivity}
\newacronym{mcc}{MCC}{Mobile Cloud Computing}
\newacronym{mchem}{MCHEM}{Massive Channel Emulator}
\newacronym{mcs}{MCS}{Modulation and Coding Scheme}
\newacronym{mec}{MEC}{Multi-access Edge Computing}
\newacronym{mec2}{MEC}{Mobile Edge Cloud}
\newacronym{mfc}{MFC}{Mobile Fog Computing}
\newacronym{mgen}{MGEN}{Multi-Generator}
\newacronym{mi}{MI}{Mutual Information}
\newacronym{mib}{MIB}{Master Information Block}
\newacronym{miesm}{MIESM}{Mutual Information Based Effective SINR}
\newacronym{mimo}{MIMO}{Multiple Input, Multiple Output}
\newacronym{ml}{ML}{Machine Learning}
\newacronym{mlr}{MLR}{Maximum-local-rate}
\newacronym[plural=\gls{mme}s,firstplural=Mobility Management Entities (MMEs)]{mme}{MME}{Mobility Management Entity}
\newacronym{mmtc}{mMTC}{Massive Machine-Type Communications}
\newacronym{mmwave}{mmWave}{millimeter wave}
\newacronym{mpdccp}{MP-DCCP}{Multipath Datagram Congestion Control Protocol}
\newacronym{mptcp}{MPTCP}{Multipath TCP}
\newacronym{mr}{MR}{Maximum Rate}
\newacronym{mrdc}{MR-DC}{Multi \gls{rat} \gls{dc}}
\newacronym{mse}{MSE}{Mean Square Error}
\newacronym{mss}{MSS}{Maximum Segment Size}
\newacronym{mt}{MT}{Mobile Termination}
\newacronym{mtd}{MTD}{Machine-Type Device}
\newacronym{mtu}{MTU}{Maximum Transmission Unit}
\newacronym{mumimo}{MU-MIMO}{Multi-user \gls{mimo}}
\newacronym{mvno}{MVNO}{Mobile Virtual Network Operator}
\newacronym{nalu}{NALU}{Network Abstraction Layer Unit}
\newacronym{nas}{NAS}{Network Attached Storage}
\newacronym{nat}{NAT}{Network Address Translation}
\newacronym{nbiot}{NB-IoT}{Narrow Band IoT}
\newacronym{nfv}{NFV}{Network Function Virtualization}
\newacronym{nfvi}{NFVI}{Network Function Virtualization Infrastructure}
\newacronym{ni}{NI}{Network Interfaces}
\newacronym{llm}{LLM}{Large Language Model}
\newacronym{dpdk}{DPDK}{Data Plane Development Kit}
\newacronym{cicd}{CI/CD}{Continuous Integration and Continuous Delivery/Deployment}
\newacronym{nic}{NIC}{Network Interface Card}
\newacronym{nlos}{NLOS}{Non-Line-of-Sight}
\newacronym{now}{NOW}{Non Overlapping Window}
\newacronym{nsm}{NSM}{Network Service Mesh}
\newacronym[type=hidden]{nr}{NR}{New Radio}
\newacronym[type=hidden]{ota}{OTA}{Over-the-Air}
\newacronym{nrf}{NRF}{Network Repository Function}
\newacronym{nsa}{NSA}{Non Stand Alone}
\newacronym{nse}{NSE}{Network Slicing Engine}
\newacronym{nssf}{NSSF}{Network Slice Selection Function}
\newacronym{o2i}{O2I}{Outdoor to Indoor}
\newacronym{oai}{OAI}{OpenAirInterface}
\newacronym{oaicn}{OAI-CN}{\gls{oai} \acrlong{cn}}
\newacronym{oairan}{OAI-RAN}{\acrlong{oai} \acrlong{ran}}
\newacronym{oam}{OAM}{Operations, Administration and Maintenance}
\newacronym{ofdm}{OFDM}{Orthogonal Frequency Division Multiplexing}
\newacronym{olia}{OLIA}{Opportunistic Linked Increase Algorithm}
\newacronym{omec}{OMEC}{Open Mobile Evolved Core}
\newacronym{onap}{ONAP}{Open Network Automation Platform}
\newacronym{onf}{ONF}{Open Networking Foundation}
\newacronym{onos}{ONOS}{Open Networking Operating System}
\newacronym{oom}{OOM}{\gls{onap} Operations Manager}
\newacronym{opnfv}{OPNFV}{Open Platform for \gls{nfv}}
\newacronym{oran}{O-RAN}{Open RAN}
\newacronym{orbit}{ORBIT}{Open-Access Research Testbed for Next-Generation Wireless Networks}
\newacronym{os}{OS}{Operating System}
\newacronym{oss}{OSS}{Operations Support System}
\newacronym{pa}{PA}{Position-aware}
\newacronym{pase}{PASE}{Prioritization, Arbitration, and Self-adjusting Endpoints}
\newacronym{pawr}{PAWR}{Platforms for Advanced Wireless Research}
\newacronym{pbch}{PBCH}{Physical Broadcast Channel}
\newacronym{pcef}{PCEF}{Policy and Charging Enforcement Function}
\newacronym{pcfich}{PCFICH}{Physical Control Format Indicator Channel}
\newacronym{pcrf}{PCRF}{Policy and Charging Rules Function}
\newacronym{pdcch}{PDCCH}{Physical Downlink Control Channel}
\newacronym{pdcp}{PDCP}{Packet Data Convergence Protocol}
\newacronym{pdf}{PDF}{Probability Density Function}
\newacronym{pdsch}{PDSCH}{Physical Downlink Shared Channel}
\newacronym{pdu}{PDU}{Packet Data Unit}
\newacronym{pgw}{PGW}{Packet Gateway}
\newacronym{phich}{PHICH}{Physical Hybrid ARQ Indicator Channel}
\newacronym{phy}{PHY}{Physical}
\newacronym{pmch}{PMCH}{Physical Multicast Channel}
\newacronym{pmi}{PMI}{Precoding Matrix Indicators}
\newacronym{powder}{POWDER}{Platform for Open Wireless Data-driven Experimental Research}
\newacronym{ppo}{PPO}{Proximal Policy Optimization}
\newacronym{ppp}{PPP}{Poisson Point Process}
\newacronym{prach}{PRACH}{Physical Random Access Channel}
\newacronym{prb}{PRB}{Physical Resource Block}
\newacronym{psnr}{PSNR}{Peak Signal to Noise Ratio}
\newacronym{pss}{PSS}{Primary Synchronization Signal}
\newacronym{pucch}{PUCCH}{Physical Uplink Control Channel}
\newacronym{pusch}{PUSCH}{Physical Uplink Shared Channel}
\newacronym{qam}{QAM}{Quadrature Amplitude Modulation}
\newacronym{qci}{QCI}{\gls{qos} Class Identifier}
\newacronym{qoe}{QoE}{Quality of Experience}
\newacronym{qos}{QoS}{Quality of Service}
\newacronym{quic}{QUIC}{Quick UDP Internet Connections}
\newacronym{rach}{RACH}{Random Access Channel}
\newacronym{ran}{RAN}{Radio Access Network}
\newacronym[firstplural=Radio Access Technologies (RATs)]{rat}{RAT}{Radio Access Technology}
\newacronym{rbg}{RBG}{Resource Block Group}
\newacronym{rcn}{RCN}{Research Coordination Network}
\newacronym{rc}{RC}{RAN Control}
\newacronym{rec}{REC}{Radio Edge Cloud}
\newacronym{red}{RED}{Random Early Detection}
\newacronym{renew}{RENEW}{Reconfigurable Eco-system for Next-generation End-to-end Wireless}
\newacronym{rf}{RF}{Radio Frequency}
\newacronym{rfc}{RFC}{Request for Comments}
\newacronym{rfr}{RFR}{Random Forest Regressor}
\newacronym{ric}{RIC}{RAN Intelligent Controller}
\newacronym{nrric}{Near-RT RIC}{Near-Real-Time \gls{ran} Intelligent Controller}
\newacronym{rlc}{RLC}{Radio Link Control}
\newacronym{rlf}{RLF}{Radio Link Failure}
\newacronym{rlnc}{RLNC}{Random Linear Network Coding}
\newacronym{rmr}{RMR}{RIC Message Router}
\newacronym{rmse}{RMSE}{Root Mean Squared Error}
\newacronym{rnis}{RNIS}{Radio Network Information Service}
\newacronym{rrc}{RRC}{Radio Resource Control}
\newacronym{rrm}{RRM}{Radio Resource Management}
\newacronym{rru}{RRU}{Remote Radio Unit}
\newacronym{rs}{RS}{Remote Server}
\newacronym{rsrp}{RSRP}{Reference Signal Received Power}
\newacronym{rsrq}{RSRQ}{Reference Signal Received Quality}
\newacronym{rss}{RSS}{Received Signal Strength}
\newacronym{rssi}{RSSI}{Received Signal Strength Indicator}
\newacronym{rtt}{RTT}{Round Trip Time}
\newacronym{ru}{RU}{Radio Unit}
\newacronym{rw}{RW}{Receive Window}
\newacronym{rx}{RX}{Receiver}
\newacronym{s1ap}{S1AP}{S1 Application Protocol}
\newacronym{sa}{SA}{standalone}
\newacronym{sack}{SACK}{Selective Acknowledgment}
\newacronym{sap}{SAP}{Service Access Point}
\newacronym{sc2}{SC2}{Spectrum Collaboration Challenge}
\newacronym{scef}{SCEF}{Service Capability Exposure Function}
\newacronym{sch}{SCH}{Secondary Cell Handover}
\newacronym{scoot}{SCOOT}{Split Cycle Offset Optimization Technique}
\newacronym{sctp}{SCTP}{Stream Control Transmission Protocol}
\newacronym{sdap}{SDAP}{Service Data Adaptation Protocol}
\newacronym{sdk}{SDK}{Software Development Kit}
\newacronym{sdm}{SDM}{Space Division Multiplexing}
\newacronym{sdma}{SDMA}{Spatial Division Multiple Access}
\newacronym{sdl}{SDL}{Shared Data Layer}
\newacronym{sdn}{SDN}{Software-defined Networking}
\newacronym{sdr}{SDR}{Software-defined Radio}
\newacronym{seba}{SEBA}{SDN-Enabled Broadband Access}
\newacronym{sgsn}{SGSN}{Serving GPRS Support Node}
\newacronym{sgw}{SGW}{Service Gateway}
\newacronym{si}{SI}{Study Item}
\newacronym{sib}{SIB}{Secondary Information Block}
\newacronym{sinr}{SINR}{Signal to Interference plus Noise Ratio}
\newacronym{sip}{SIP}{Session Initiation Protocol}
\newacronym{siso}{SISO}{Single Input, Single Output}
\newacronym{sla}{SLA}{Service Level Agreement}
\newacronym{sm}{SM}{Service Model}
\newacronym{e2sm}{E2SM}{E2 Service Model}
\newacronym{e2ap}{E2AP}{E2 Application Protocol}
\newacronym{smf}{SMF}{Session Management Function}
\newacronym{smo}{SMO}{Service Management and Orchestration}
\newacronym{sms}{SMS}{Short Message Service}
\newacronym{smsgmsc}{SMS-GMSC}{\gls{sms}-Gateway}
\newacronym{snr}{SNR}{Signal-to-Noise-Ratio}
\newacronym{son}{SON}{Self-Organizing Network}
\newacronym{sptcp}{SPTCP}{Single Path TCP}
\newacronym{srb}{SRB}{Service Radio Bearer}
\newacronym{srn}{SRN}{Standard Radio Node}
\newacronym{srs}{SRS}{Sounding Reference Signal}
\newacronym{ss}{SS}{Synchronization Signal}
\newacronym{sss}{SSS}{Secondary Synchronization Signal}
\newacronym{st}{ST}{Spanning Tree}
\newacronym{svc}{SVC}{Scalable Video Coding}
\newacronym{tb}{TB}{Transport Block}
\newacronym{tcp}{TCP}{Transmission Control Protocol}
\newacronym{tdd}{TDD}{Time Division Duplexing}
\newacronym{tdm}{TDM}{Time Division Multiplexing}
\newacronym{tdma}{TDMA}{Time Division Multiple Access}
\newacronym{tfl}{TfL}{Transport for London}
\newacronym{tfrc}{TFRC}{TCP-Friendly Rate Control}
\newacronym{tft}{TFT}{Traffic Flow Template}
\newacronym{tgen}{TGEN}{Traffic Generator}
\newacronym{tip}{TIP}{Telecom Infra Project}
\newacronym{tm}{TM}{Transparent Mode}
\newacronym{to}{TO}{Telco Operator}
\newacronym{tr}{TR}{Technical Report}
\newacronym{trp}{TRP}{Transmitter Receiver Pair}
\newacronym{ts}{TS}{Technical Specification}
\newacronym{tti}{TTI}{Transmission Time Interval}
\newacronym{ttt}{TTT}{Time-to-Trigger}
\newacronym{tx}{TX}{Transmitter}
\newacronym{uas}{UAS}{Unmanned Aerial System}
\newacronym{uav}{UAV}{Unmanned Aerial Vehicle}
\newacronym{udm}{UDM}{Unified Data Management}
\newacronym{udp}{UDP}{User Datagram Protocol}
\newacronym{udr}{UDR}{Unified Data Repository}
\newacronym{ue}{UE}{User Equipment}
\newacronym{uhd}{UHD}{\gls{usrp} Hardware Driver}
\newacronym{ul}{UL}{Uplink}
\newacronym{um}{UM}{Unacknowledged Mode}
\newacronym{uml}{UML}{Unified Modeling Language}
\newacronym{upa}{UPA}{Uniform Planar Array}
\newacronym{upf}{UPF}{User Plane Function}
\newacronym{urllc}{URLLC}{Ultra Reliable and Low Latency Communications}
\newacronym{usa}{U.S.}{United States}
\newacronym{usim}{USIM}{Universal Subscriber Identity Module}
\newacronym{usrp}{USRP}{Universal Software Radio Peripheral}
\newacronym{utc}{UTC}{Urban Traffic Control}
\newacronym{vim}{VIM}{Virtualization Infrastructure Manager}
\newacronym{vm}{VM}{Virtual Machine}
\newacronym{vnf}{VNF}{Virtual Network Function}
\newacronym{volte}{VoLTE}{Voice over \gls{lte}}
\newacronym{voltha}{VOLTHA}{Virtual OLT HArdware Abstraction}
\newacronym{vr}{VR}{Virtual Reality}
\newacronym{vran}{vRAN}{Virtualized RAN}
\newacronym{vss}{VSS}{Video Streaming Server}
\newacronym{wbf}{WBF}{Wired Bias Function}
\newacronym{wf}{WF}{Waterfilling}
\newacronym{wg}{WG}{Working Group}
\newacronym{wlan}{WLAN}{Wireless Local Area Network}
\newacronym{osm}{OSM}{Open Source \gls{nfv} Management and Orchestration}
\newacronym{pnf}{PNF}{Physical Network Function}
\newacronym{drl}{DRL}{Deep Reinforcement Learning}
\newacronym{rl}{RL}{Reinforcement Learning}
\newacronym{fpv}{FPV}{First-Person View}
\newacronym{isr}{ISR}{Intelligence, Surveillance, and Reconnaissance}
\newacronym{cg}{CG}{Configured Grant}
\newacronym{mtc}{MTC}{Machine-type Communications}
\newacronym{osc}{OSC}{O-RAN Software Community}
\newacronym{mns}{MnS}{Management Services}
\newacronym{ves}{VES}{\gls{vnf} Event Stream}
\newacronym{ei}{EI}{Enrichment Information}
\newacronym{fh}{FH}{Fronthaul}
\newacronym{fft}{FFT}{Fast Fourier Transform}
\newacronym{laa}{LAA}{Licensed-Assisted Access}
\newacronym{plfs}{PLFS}{Physical Layer Frequency Signals}
\newacronym{ptp}{PTP}{Precision Time Protocol}
\newacronym{ntp}{NTP}{Network Time Protocol}
\newacronym{cbrs}{CBRS}{Citizen Broadband Radio Service}
\newacronym{rnti}{RNTI}{Radio Network Temporary Identifier}
\newacronym{tbs}{TBS}{Transport Block Size}
\newacronym{nfd}{NFD}{Node Feature Discovery}
\newacronym{mcp}{MCP}{Model Context Protocol}
\newacronym{vpn}{VPN}{Virtual Private Network}
\newacronym{onr}{ONR}{Office of Naval Research}
\newacronym{afosr}{AFOSR}{Air Force Office of Scientific Research}
\newacronym{afrl}{AFRL}{Air Force Research Laboratory}
\newacronym{arl}{ARL}{Army Research Laboratory}

\newacronym{arc}{ARC}{Aerial Research Cloud}

\newacronym{ct}{CT}{Continuous Testing}
\newacronym{mno}{MNO}{Mobile Network Operator}
\newacronym{oci}{OCI}{Open Container Initiative}
\newacronym{macsec}{MACsec}{Media Access Control Security}
\newacronym{pt}{PT}{Plain Text}
\newacronym{cuda}{CUDA}{Compute Unified Device Architecture}
\newacronym{dsp}{DSP}{Digital Signal Processing}

\newacronym{cus}{CUS}{Control, User, Synchronization}
\newacronym{dpd}{DPD}{Digital Pre-Distorsion}
\newacronym{cfr}{CFR}{Crest Factor Reduction}
\newacronym{pci}{PCIe}{Peripheral Component Interconnect Express}
\newacronym{dpu}{DPU}{Data Processing Unit}
\newacronym{rfsoc}{RFSoC}{Radio Frequency System-on-Chip}
\newacronym{if}{IF}{Intermediate Frequency}
\newacronym{nyu}{NYU}{New York University}
\newacronym{gh}{GH}{Grace Hopper}
\newacronym{trl}{TRL}{Technology Readiness Level}
\newacronym{srfa}{SRFA}{Special Research Focus Area}
\newacronym{qsfp}{QSFP}{quad small form factor pluggable}
\newacronym{pse}{PSE}{Performance Specialized Engine}
\newacronym{cae}{CAE}{Cognitive Analysis Engine}
\newacronym{simd}{SIMD}{Single Instruction/Multiple Data}
\newacronym{rt}{RT}{Real-Time}
\newacronym{nrt}{NRT}{Non-Real-Time}
\newacronym{asm}{ASM}{Advanced Sleep Mode}
\newacronym{aoa}{AoA}{Angle of Arrival}
\newacronym{eaxcid}{eAxC\_ID}{extended Antenna-Carrier Identifier}
\newacronym{bwp}{BWP}{Bandwidth Part}
\newacronym{dfe}{DFE}{Digital Front-End}
\newacronym{spi}{SPI}{Serial Peripheral Interface}
\newacronym{gpio}{GPIO}{General Purpose Input/Output}
\newacronym{nco}{NCO}{Numerically Controlled Oscillator}
\newacronym{lo}{LO}{Local Oscillator}
\newacronym{lna}{LNA}{Low-Noise Amplifier}
\newacronym{pll}{PLL}{Phased-Locked Loop}
\newacronym{som}{SOM}{System-on-Module}
\newacronym{papr}{PAPR}{Peak-to-Average Power Ratio}
\newacronym{pcb}{PCB}{Printed Circuit Board}
\newacronym{gcpw}{GCPW}{Grounded Co-Planar Waveguide}
\newacronym{cnn}{CNN}{Convolutional Neural Network}
\newacronym{gmp}{GMP}{Generalized Memory Polynomial}
\newacronym{ngrg}{nGRG}{next Generation Research Group}
\newacronym{mrl}{MRL}{Manufacturing Readiness Level}
\newacronym{fr}{FR}{Frequency Range}

\newacronym{sbom}{SBOM}{Software Bill of Materials}
\newacronym{hbom}{HBOM}{Hardware Bill of Materials}
\newacronym{vex}{VEX}{Vulnerability Exploitability eXchange}
\newacronym{dos}{DoS}{Denial of Service}
\newacronym{sme}{SME}{Small-Medium Enterprise}

\newacronym{ulpi}{ULPI}{Uplink Performance Improvement}
\newacronym{oem}{OEM}{Original Equipment Manufacturer}
\newacronym{nsin}{NSIN}{National Security Innovation Network}
\newacronym{dod}{DoD}{Department of Defense}
\newacronym{arpu}{ARPU}{Average Revenue per User}
\newacronym{opex}{OPEX}{operational expenses}
\newacronym{txb}{TXB}{Transmit Beam}
\newacronym{cve}{CVE}{Common Vulnerabilities and Exposure}
\newacronym{json}{JSON}{JavaScript Object Notation}
\newacronym{it}{IT}{Information Technology}
\newacronym{ci}{CI}{Continuous Integration}
\newacronym{nlp}{NLP}{Natural Language Processing}
\newacronym{nl}{NL}{Natural Language}
\newacronym{ulsch}{ULSCH}{Uplink Shared Channel}
\newacronym{phr}{PHR}{Power Headroom Report}
\newacronym{fwa}{FWA}{Fixed Wireless Access}
\newacronym{iq}{IQ}{In-Phase/Quadrature}
\newacronym{ram}{RAM}{Random Access Memory}
\newacronym{xr}{XR}{eXtended Reality}
\newacronym{aws}{AWS}{Amazon Web Services}
\newacronym{ra}{RA}{Resource Allocation}
\newacronym{pc}{PC}{Power Control}
\newacronym{a2a}{A2A}{Agent-To-Agent}
\newacronym{ssb}{SSB}{Synchronization Signal Block}
\newacronym{ta}{TA}{Timing Advance}
\newacronym{sr}{SR}{Scheduling Request}
\newacronym{v2x}{V2X}{Vehicle-To-Everything}
\newacronym{nn}{NN}{Neural Network}
\newacronym{tw}{TW}{Tolerance Window}
\newacronym{sdu}{SDU}{Service Data Unit}
\newacronym{cb}{CB}{Callback}
\usepackage{booktabs}
\usepackage{balance}
\def\BibTeX{{\rm B\kern-.05em{\sc i\kern-.025em b}\kern-.08em T\kern-.1667em\lower.7ex\hbox{E}\kern-.125emX}}

\usepackage{enumitem}
\setlist[itemize]{leftmargin=*}

\usepackage{subcaption}
\usepackage[many]{tcolorbox}
\usepackage{lipsum}

\usepackage{xcolor}

\begin{document}
\bstctlcite{IEEEexample:BSTcontrol}

\title{AUGUSTE: Online-Learning dApp for Predictive URLLC Scheduling%
}

\IEEEoverridecommandlockouts
\author{%
\IEEEauthorblockN{Maxime Elkael$^\dagger$, Michele Polese$^\dagger$, Yunseong Lee$^\ddagger$, Koichiro Furueda$^\ddagger$, Tommaso Melodia$^\dagger$}%
\thanks{$^\dagger$Institute for Intelligent Networked Systems, Northeastern University (NU), Boston, MA, USA. E-mail: \{m.elkael, m.polese, melodia\}@northeastern.edu. $^\ddagger$SoftBank Corp.\ (SB), Japan. E-mail: yunseong.lee.1992@ieee.org, koichiro.furueda@g.softbank.co.jp.}%
\thanks{This work is partially supported by OUSD(R\&E) through Army Research Laboratory Cooperative Agreement Number W911NF-24-2-0065. The views and conclusions contained in this document are those of the authors and should not be interpreted as representing the official policies, either expressed or implied, of the Army Research Laboratory or the U.S.\ Government. The U.S.\ Government is authorized to reproduce and distribute reprints for Government purposes notwithstanding any copyright notation herein. This work is also partially supported by the U.S.\ NSF under award CNS-2112471 and by SoftBank Corp.}%
}

\IEEEaftertitletext{\vspace{-1\baselineskip}}
\maketitle

\glsunset{5g}
\glsunset{3gpp}

\begin{abstract}
\gls{urllc} was one of the main motivations behind \gls{5g}, with \gls{3gpp} advertising 1--10~ms latency targets for applications such as industrial automation, \gls{v2x}, tactical edge networking, and unmanned-system control. Years on, real \gls{5g} \gls{tdd} networks still show median \gls{ul} round-trip times in the 50--70~ms range, largely because of the \gls{sr} procedure that a \gls{ue} must complete before transmitting \gls{ul} data. Existing remedies, primarily \gls{cg} scheduling, only eliminate this overhead for strictly periodic traffic and require cross-layer synchronization, which has limited their adoption. We propose \framework (\emph{Anticipatory Uplink Grants for URLLC via Self-Adapting Temporal Estimation}), a learning-based \gls{mac} scheduling framework that embeds online \gls{ml} models in the \gls{ul} scheduler to predict packet arrivals and proactively allocate resources before an \gls{sr} is issued. An adaptive state machine alternates between a \emph{learning} phase that collects unbiased arrival statistics and a \emph{confident} phase that exploits the learned predictions to schedule only when traffic is expected. We evaluate \framework on a real \gls{5g} testbed running OpenAirInterface across three \gls{urllc} traffic patterns (request-response, \gls{ml} edge inference, and periodic autonomous reporting), and show that it operates at the best achievable point on the latency--overhead trade-off: it matches always-on scheduling's median \gls{rtt} ($\sim 10$~ms, halving the $\sim 20$~ms \gls{sr}-based baseline) at roughly one-tenth its resource cost ($7$--$10\%$ overhead).
\end{abstract}
\vspace*{-5pt}
\section{Introduction}

Recent counter-\gls{uav} operations have made low-latency tactical wireless networking a frontline requirement. \gls{fpv} \glspl{uav} have become a defining feature of contemporary drone warfare, with teleoperation requiring tight latency and jitter bounds to maintain control and detection of fast-moving aerial objects. Further, real-time \gls{ai} inference is driving low latency requirements for reporting from tactical edge sensors and \glspl{uav}. Finally, the U.S.\ \gls{dod} Replicator initiative aims to field thousands of attritable (i.e., expendable) autonomous systems~\cite{replicator2024crs}. We argue their effectiveness will depend on the network supporting mass, bursty, low-latency \gls{ul} communications. These tactical demands fall in the \gls{urllc} service class (sub-millisecond to tens-of-millisecond latencies at $\geq 99.999\%$ reliability, depending on use case~\cite{ts22261,popovski2018wireless}), and the same budget is load-bearing for applications such as \gls{v2x}, telemedicine with haptic feedback, and industrial automation. In practice, however, real \gls{5g} \gls{tdd} deployments fall short on both fronts: large-scale measurements of commercial \gls{5g} \gls{sa} networks~\cite{ghoshal2025first} report median \gls{rtt}s in the $50$--$70$~ms range with tails reaching $200$~ms.

\begin{figure}[t]
    \centering
    {\setlength{\fboxsep}{1pt}\framebox[\linewidth][c]{\ref{cdflegend}}}\\[3pt]
    \begin{minipage}[t]{0.5\linewidth}
        \centering
        \includegraphics[width=\linewidth]{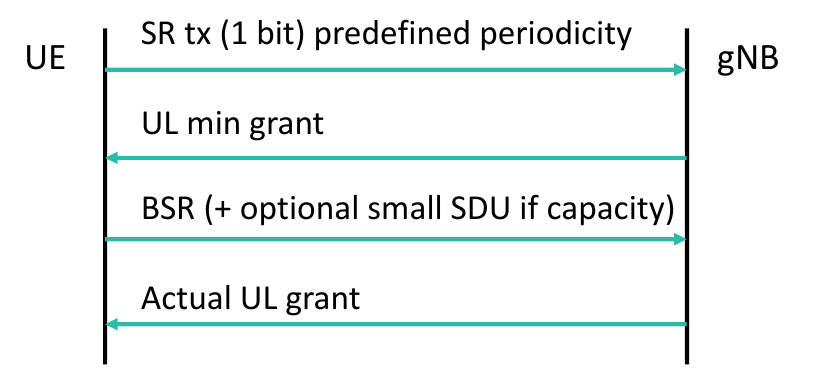}
    \end{minipage}%
    \begin{minipage}[t]{0.5\linewidth}
        \centering
        \vspace{-50pt}
\begin{tikzpicture}[baseline=(current bounding box.north), trim axis left, trim axis right]

\definecolor{darkgrey176}{RGB}{176,176,176}
\definecolor{lightgrey204}{RGB}{204,204,204}

\begin{axis}[
width=1.2\linewidth,
height=0.6\linewidth,
grid=major,
legend cell align={left},
legend style={
  draw=none,
  fill opacity=0,
  text opacity=1,
  font=\scriptsize
},
legend columns=2,
legend to name=cdflegend,
thick,
tick align=outside,
tick pos=left,
x grid style={darkgrey176},
xlabel={RTT (ms)},
xlabel style={font=\scriptsize},
xticklabel style={font=\scriptsize},
xmajorgrids,
xmin=0, xmax=40,
xtick style={color=black},
y grid style={darkgrey176},
yticklabel style={font=\scriptsize},
ymajorgrids,
ymin=0, ymax=1,
ytick={0,1},
ytick style={color=black}
]
\addplot [thick, blue]
table {%
6.88 0.001
6.92 0.014
6.94 0.028
6.96 0.042
7.38 0.056
7.88 0.07
7.9 0.084
7.91 0.098
7.93 0.111
7.99 0.125
8.36 0.139
8.4 0.153
8.91 0.167
8.93 0.181
8.95 0.195
9.34 0.209
9.37 0.222
9.39 0.236
9.4 0.25
9.44 0.264
15 0.278
15.5 0.292
15.9 0.306
16 0.32
16.9 0.333
17.1 0.347
17.9 0.361
18.1 0.375
18.9 0.389
19.9 0.403
19.9 0.417
19.9 0.431
19.9 0.444
19.9 0.458
19.9 0.472
20 0.486
20.1 0.5
20.4 0.514
20.5 0.528
20.9 0.542
20.9 0.555
20.9 0.569
20.9 0.583
20.9 0.597
21 0.611
21.4 0.625
21.9 0.639
21.9 0.653
21.9 0.666
21.9 0.68
21.9 0.694
22.7 0.708
22.9 0.722
22.9 0.736
22.9 0.75
22.9 0.764
23 0.777
23.6 0.791
23.9 0.805
23.9 0.819
23.9 0.833
23.9 0.847
24 0.861
24.8 0.875
24.9 0.888
24.9 0.902
24.9 0.916
25.8 0.93
25.9 0.944
26.9 0.958
28.4 0.972
28.9 0.986
32.9 0.999
33.5 1
};
\addlegendentry{Default SR-based scheduling}
\addplot [thick, red, dashed]
table {%
6.4 0.001
6.9 0.014
6.91 0.028
6.92 0.042
6.93 0.056
6.94 0.07
7.37 0.084
7.41 0.098
7.43 0.111
7.47 0.125
7.87 0.139
7.88 0.153
7.89 0.167
7.9 0.181
7.9 0.195
7.91 0.209
7.91 0.222
7.92 0.236
7.92 0.25
7.93 0.264
7.94 0.278
7.95 0.292
8.44 0.306
8.89 0.32
8.9 0.333
8.9 0.347
8.91 0.361
8.92 0.375
8.92 0.389
8.93 0.403
8.94 0.417
8.96 0.431
9.86 0.444
9.88 0.458
9.89 0.472
9.9 0.486
9.9 0.5
9.9 0.514
9.91 0.528
9.91 0.542
9.92 0.555
9.92 0.569
9.93 0.583
9.93 0.597
9.94 0.611
9.98 0.625
10.9 0.639
10.9 0.653
10.9 0.666
10.9 0.68
10.9 0.694
10.9 0.708
10.9 0.722
10.9 0.736
10.9 0.75
10.9 0.764
10.9 0.777
10.9 0.791
11.2 0.805
11.4 0.819
11.4 0.833
11.9 0.847
11.9 0.861
11.9 0.875
11.9 0.888
11.9 0.902
13.9 0.916
13.9 0.93
13.9 0.944
13.9 0.958
15.7 0.972
16 0.986
18.9 0.999
18.9 1
};
\addlegendentry{Always scheduling in UL}
\end{axis}

\end{tikzpicture}
    \end{minipage}\\[0.3em]
    \begin{minipage}[t]{0.5\linewidth}
        \centering
        \captionof{figure}{\gls{sr} procedure.}
        \label{fig:schedReq}
    \end{minipage}%
    \begin{minipage}[t]{0.5\linewidth}
        \centering
        \captionof{figure}{\gls{rtt} w/ \& w/o always-on \gls{ul}.}
        \label{fig:sr-vs-always}
    \end{minipage}
    \vspace{-25pt}
\end{figure}

The main culprit on the \gls{ul} is the \gls{sr} procedure. As shown in Fig.~\ref{fig:schedReq}, a \gls{ue} with new \gls{ul} data must first signal its intent to transmit by sending a \gls{sr} on a periodically configured \gls{pucch} opportunity; the \gls{gnb} then allocates a small grant, receives the \gls{bsr}, and finally schedules the actual data on a subsequent \gls{ul} opportunity. For sparse traffic, this has to be done for every burst, with total overhead routinely exceeding 15~ms. The overhead is worst in \gls{tdd} due to slot-alignment delays, hence, in this paper, we focus on \gls{tdd}, but note the same \gls{sr} procedure also applies in FDD.

\glsreset{cg}

To reduce this \gls{sr} overhead, \gls{3gpp} introduced \gls{cg} (or grant-free) scheduling~\cite{ts38321,le2021urllc}, in which the \gls{gnb} pre-allocates periodic \gls{ul} grants without a \gls{sr}. This (i) only fits strictly periodic traffic, which is a poor match for event-driven (e.g., tactical sensing) or multi-source (e.g., a robot with multiple sensors) traffic, and (ii) imposes tight cross-layer synchronization between the Application and RRC Layer at the \gls{ue}, complicating deployment and limiting adoption.

Beyond \gls{cg}, ML-based scheduling has been studied at the \gls{mac} layer for \gls{dl} resource allocation~\cite{altam2020leasch} and for \gls{urllc}/\gls{embb} minislot coexistence~\cite{anand2020joint}, and with O-RAN xApps for \gls{urllc} orchestration~\cite{sohaib2024drl}. Cellular traffic prediction has also been studied extensively, but the bulk of this literature operates at coarser timescales (minutes to hours) and at cell-aggregate granularity~\cite{jiang2022cellular}. \framework targets a different regime: per-slot \gls{ul} grant decisions that bypass the \gls{sr} procedure, via the \textit{dApp} paradigm~\cite{lacava2025dapps}, i.e., with control logic deployed directly inside the \gls{ran} node.

A simple workaround to reduce latency is to schedule \gls{urllc} \glspl{ue} proactively (i.e. without a \gls{sr} or \gls{cg}) in every slot (\emph{always-on} scheduling), recently demonstrated in srsRAN as part of a broader latency-reduction effort~\cite{gong2025towardsurllc}. This eliminates \gls{sr} delay but wastes over 90\% of the allocated radio resources for sparse traffic, while having poor scaling. \framework instead schedules \emph{proactively but selectively}: it predicts when \gls{ul} packets are about to arrive and pre-allocates resources only at those times. The predictor is trained \emph{online}: its training data, the true \gls{ul} arrival times, is generated by the framework's own actions rather than by an offline annotation pipeline. Because the \gls{sr} mechanism biases reactive measurements, the framework cannot learn from passive observation; the proactive grants issued during a brief learning phase are what surface the un-distorted arrival times, which the predictor then consumes as training data. No offline dataset or operator annotation is required.

\textbf{Our Contributions.} (i) The first \gls{ul} scheduling dApp that learns online from multiple in-\gls{ran} events (\gls{ul} receptions, \gls{dl} transmissions, and proactive-grant failures) to predict packet arrivals and bypass the \gls{sr} procedure on a real \gls{5g} \gls{sa} platform. (ii) An adaptive three-state (Learning/Confident/Idle) supervisor that trades off radio overhead and latency without \gls{rrc} reconfiguration or prior knowledge of the traffic distribution, parameterized by two knobs 
that expose the latency-overhead trade-off explicitly. (iii) An end-to-end implementation of \framework as a custom \gls{oai} scheduler dApp, evaluated empirically on a real \gls{5g} \gls{sa} testbed across three representative \gls{urllc} traffic patterns. We measure a 50\% median \gls{rtt} reduction (20~ms to 10~ms) at 7--10\% radio overhead, and show that \framework operates at the best achievable point on this trade-off.

The rest of the paper is organized as follows. Section~\ref{sec:problem} formalizes the latency--overhead trade-off; Section~\ref{sec:solution} describes the proposed framework; Section~\ref{sec:setup} details the experimental setup; Section~\ref{sec:results} presents results; and Section~\ref{sec:conclusion} concludes the article.

\section{Problem Formulation}
\label{sec:problem}

The \gls{ul} scheduling problem in \gls{5g} \gls{urllc} is a trade-off between latency and resource efficiency. This section quantifies the \gls{sr} overhead, formalizes the multi-objective trade-off, and identifies the resulting requirements for an effective solution.

\subsection{Quantifying the Scheduling Request Overhead}

Beyond the qualitative picture of Fig.~\ref{fig:schedReq}, the \gls{sr} procedure contributes several compounding delays. The \gls{ue} first waits for its next \gls{sr} opportunity. This comes with a delay ranging from 0 to the configured \gls{sr} periodicity, typically 10--40~ms and up to 640~ms in \gls{5g}. Limited \gls{pucch} resources force operators to use sparser periodicities as the number of \glspl{ue} grows. Then, the \gls{sr} is followed by a \gls{dci} grant on the next \gls{dl} opportunity and \gls{ul} transmission on the next \gls{ul} slot (after at least one full \gls{tdd} period). In this, a small initial grant typically carries only the \gls{bsr}, requiring yet another \gls{tdd} period for the actual data. Even in the best case, the total overhead is at least the \gls{sr} periodicity plus one \gls{tdd} period, and in practice ranges from 15--50~ms, far exceeding the 1--10~ms \gls{urllc} targets. The \gls{sr} procedure is also a major source of jitter: the wait time depends on when a packet arrives relative to the \gls{sr} periodicity, producing a near-uniform random delay that is damaging for applications demanding deterministic latency bounds.

Fig.~\ref{fig:sr-vs-always} quantifies this overhead on our \gls{oai} testbed under no competing load (representative of a lightly-loaded cell or a \gls{urllc} slice with reserved resources). The default \gls{sr}-based scheduling exhibits \gls{rtt} values in the 15--25~ms range with median $\approx$~20~ms and significant variance: already below commercial measurements but still well above \gls{urllc} targets. In contrast, when the \gls{ul} scheduler systematically allocates resources without waiting for an \gls{sr} (``always scheduling''), the \gls{rtt} drops to 7--12~ms with a much tighter distribution. This $\approx$50\% latency reduction directly corresponds to eliminating the \gls{sr} overhead and its associated variability.

\subsection{The Resource Efficiency Dilemma}

While always scheduling successfully reduces latency and jitter, it creates an unsustainable resource overhead. For sparse \gls{urllc} traffic patterns typical in industrial \gls{iot} (e.g., periodic sensor reports at 10-100Hz), this approach wastes over 90\% of allocated resources. With $N$ \gls{urllc} \glspl{ue}, the system would require $N$ times the resources, severely limiting scalability.

We formulate this as the multi-objective optimization problem \eqref{eq:multiobj}, seeking Pareto-optimal scheduling strategies that trade off three objectives: worst-case expected latency, worst-case jitter, and total slice resource consumption.
\begin{equation}\label{eq:multiobj}\tag{P-URLLC}
\begin{aligned}
\underset{S \in \mathcal{S}}{\text{minimize}} \quad & \begin{bmatrix}
\max_{i \in \mathcal{N}} \mathbb{E}[L_{sched}^i(S)] \\
\max_{i \in \mathcal{N}} V(L_{sched}^i(S)) \\
\sum_{i \in \mathcal{N}} U_i(S)
\end{bmatrix} \\
\text{subject to} \quad & U_i(S) = \frac{\mathbb{E}[R_i(S)]}{R_{slice}/|\mathcal{N}|}, \quad \forall i \in \mathcal{N},
\end{aligned}
\end{equation}
where:
\begin{itemize}[leftmargin=*]
\item $\mathcal{N}$ is the set of \glspl{ue} in a \gls{urllc} slice, assumed to have homogeneous \gls{qos} requirements
\item $S \in \mathcal{S}$ is a feasible scheduling strategy
\item $L_{sched}^i(S)$ is the scheduling-induced latency for \gls{ue} $i$
\item $\mathbb{E}[L_{sched}^i(S)]$ is the expected scheduling latency for \gls{ue} $i$
\item $V(L_{sched}^i(S))$ is the latency variance (jitter) for \gls{ue} $i$
\item $U_i(S)$ is the resource consumption ratio for \gls{ue} $i$
\item $R_i(S)$ is the random variable representing instantaneous resources allocated to \gls{ue} $i$
\item $\mathbb{E}[R_i(S)]$ is the time-averaged resources allocated to \gls{ue} $i$
\item $R_{slice}$ is the total \gls{prb} resources allocated to this \gls{urllc} slice
\item $|\mathcal{N}|$ is the number of \glspl{ue} in the slice
\end{itemize}
The three objectives reflect different aspects of \gls{urllc} requirements within a slice: the first ensures no individual \gls{ue} experiences excessive expected latency (fairness and worst-case guarantee), the second ensures predictable performance by minimizing the worst-case jitter across all \glspl{ue}, and the third minimizes total slice resource consumption for scalability. The expectations and variances are taken over the stochastic packet arrival process and channel variations.

This problem is challenging because $L_{sched}^i(S)$ and $U_i(S)$ are inversely coupled: strategies that minimize latency (e.g., always-on scheduling) maximize resource usage, while strategies that minimize overhead (e.g., \gls{sr}-based scheduling) maximize latency. The discrete nature of \gls{prb} allocation and the stochastic arrival patterns of traffic make~\eqref{eq:multiobj} non-convex, with no closed-form solution. Standard approaches select fixed points on the trade-off curve: \gls{sr}-based scheduling chooses minimum resources at the cost of high latency, always-on scheduling chooses minimum latency at the cost of maximum resources, and \gls{cg} scheduling attempts a middle ground for the restricted case of strictly periodic traffic. None of these can adapt their operating point to the actual arrival process, nor can they exploit the temporal correlations in \gls{urllc} traffic to improve both objectives simultaneously.

\section{System Design}
\label{sec:solution}

Solving~\eqref{eq:multiobj} exactly is intractable: the latency random variable $L^i_{sched}(S)$ depends on the unknown \gls{ul} arrival process, and no closed-form Pareto frontier exists. We instead propose an online strategy $S_{\text{ML}}$ that (i) estimates the arrival process from in-band scheduler events using lightweight \gls{ml} models embedded directly in the \gls{mac} scheduler as a Layer-2 dApp, and (ii) lets the operator tune its position on the latency-overhead trade-off through two knobs, a tolerance window $\mathrm{TW}$ and a slot restriction $N$, introduced in Section~\ref{sec:state-machine}. Section~\ref{sec:results} shows that the learned predictions are accurate enough to push $S_{\text{ML}}$ close to simultaneous minimum overhead and minimum latency. The framework relies on an adaptive state machine that alternates between aggressive learning phases that collect unbiased arrival statistics and prediction-driven proactive scheduling phases that exploit them. Before describing the architecture, we first show that even a primitive instance of $S_{\text{ML}}$ already approaches always-on latency at a fraction of its overhead.

\subsection{Proof of Concept}

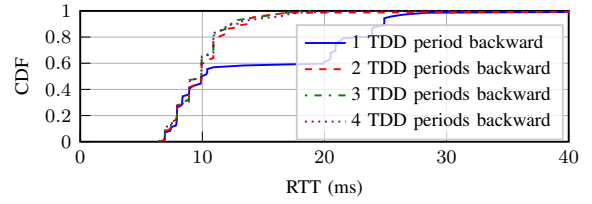
\begin{figure}
\centering
    \setlength\fwidth{0.9\linewidth}
    \setlength\fheight{2cm}
\begin{tikzpicture}[
scale=0.9
]

\definecolor{darkgrey176}{RGB}{176,176,176}
\definecolor{green}{RGB}{0,128,0}
\definecolor{lightgrey204}{RGB}{204,204,204}
\definecolor{purple}{RGB}{128,0,128}

\begin{axis}[
width=\linewidth,
height=0.4\linewidth,
grid=major,
legend cell align={left},
legend style={
  fill opacity=0.8,
  draw opacity=1,
  text opacity=1,
  at={(0.99,0.01)},
  anchor=south east,
  draw=lightgrey204,
  font=\footnotesize
},
thick,
tick align=inside,
tick pos=left,
x grid style={darkgrey176},
xlabel={RTT (ms)},
xmajorgrids,
xmin=0, xmax=40,
xtick style={color=black},
y grid style={darkgrey176},
ylabel={CDF},
ymajorgrids,
ymin=0, ymax=1,
ytick style={color=black},
xlabel style={font=\footnotesize},
ylabel style={font=\footnotesize},
ticklabel style={font=\footnotesize}
]
\addplot [thick, blue]
table {%
6.88 0.001
6.9 0.014
6.92 0.028
6.93 0.042
6.94 0.056
6.95 0.07
7.38 0.084
7.42 0.098
7.44 0.111
7.85 0.125
7.88 0.139
7.89 0.153
7.9 0.167
7.91 0.181
7.92 0.195
7.92 0.209
7.93 0.222
7.93 0.236
7.94 0.25
7.95 0.264
8.29 0.278
8.33 0.292
8.35 0.306
8.36 0.32
8.37 0.333
8.39 0.347
8.89 0.361
8.91 0.375
8.93 0.389
8.94 0.403
9.08 0.417
9.4 0.431
9.86 0.444
9.89 0.458
9.91 0.472
9.92 0.486
9.94 0.5
10.3 0.514
10.4 0.528
10.4 0.542
10.4 0.555
10.9 0.569
13.4 0.583
19.9 0.597
20 0.611
20.4 0.625
20.5 0.639
20.5 0.653
20.5 0.666
20.5 0.68
20.9 0.694
20.9 0.708
20.9 0.722
20.9 0.736
21 0.75
21.1 0.764
21.4 0.777
21.5 0.791
23.9 0.805
23.9 0.819
23.9 0.833
23.9 0.847
23.9 0.861
24 0.875
24.8 0.888
24.9 0.902
24.9 0.916
24.9 0.93
24.9 0.944
25.5 0.958
26.5 0.972
28.4 0.986
56.9 0.999
83.9 1
};
\addlegendentry{1 TDD period backward}
\addplot [thick, red, dashed]
table {%
6.46 0.001
6.9 0.014
6.91 0.028
6.92 0.042
6.93 0.056
6.95 0.07
7.04 0.084
7.39 0.098
7.4 0.111
7.42 0.125
7.44 0.139
7.47 0.153
7.88 0.167
7.89 0.181
7.9 0.195
7.9 0.209
7.91 0.222
7.92 0.236
7.92 0.25
7.93 0.264
7.94 0.278
8.4 0.292
8.45 0.306
8.89 0.32
8.9 0.333
8.9 0.347
8.91 0.361
8.91 0.375
8.92 0.389
8.93 0.403
8.94 0.417
9.32 0.431
9.86 0.444
9.89 0.458
9.89 0.472
9.9 0.486
9.91 0.5
9.91 0.514
9.91 0.528
9.92 0.542
9.92 0.555
9.93 0.569
9.94 0.583
9.95 0.597
10.4 0.611
10.4 0.625
10.9 0.639
10.9 0.653
10.9 0.666
10.9 0.68
10.9 0.694
10.9 0.708
10.9 0.722
10.9 0.736
10.9 0.75
10.9 0.764
11 0.777
11.4 0.791
11.4 0.805
11.8 0.819
11.9 0.833
11.9 0.847
11.9 0.861
12.9 0.875
12.9 0.888
13.9 0.902
13.9 0.916
13.9 0.93
13.9 0.944
14.5 0.958
15.9 0.972
16.9 0.986
57.9 0.999
61.3 1
};
\addlegendentry{2 TDD periods backward}
\addplot [thick, green, dash pattern=on 1pt off 3pt on 3pt off 3pt]
table {%
6.42 0.001
6.9 0.014
6.92 0.028
6.92 0.042
6.93 0.056
6.93 0.07
6.94 0.084
6.94 0.098
6.96 0.111
7.42 0.125
7.43 0.139
7.87 0.153
7.89 0.167
7.9 0.181
7.91 0.195
7.92 0.209
7.92 0.222
7.92 0.236
7.93 0.25
7.93 0.264
7.94 0.278
8.15 0.292
8.44 0.306
8.89 0.32
8.9 0.333
8.91 0.347
8.92 0.361
8.92 0.375
8.93 0.389
8.93 0.403
8.93 0.417
8.94 0.431
8.94 0.444
8.95 0.458
8.96 0.472
9.88 0.486
9.9 0.5
9.91 0.514
9.91 0.528
9.92 0.542
9.92 0.555
9.92 0.569
9.93 0.583
9.93 0.597
9.93 0.611
9.94 0.625
9.94 0.639
9.95 0.653
10.7 0.666
10.9 0.68
10.9 0.694
10.9 0.708
10.9 0.722
10.9 0.736
10.9 0.75
10.9 0.764
10.9 0.777
10.9 0.791
10.9 0.805
11 0.819
11 0.833
11.4 0.847
11.9 0.861
11.9 0.875
11.9 0.888
12.4 0.902
12.9 0.916
12.9 0.93
13.9 0.944
14.9 0.958
15.9 0.972
16.9 0.986
18 0.999
18.9 1
};
\addlegendentry{3 TDD periods backward}
\addplot [thick, purple, dotted]
table {%
6.41 0.001
6.9 0.014
6.91 0.028
6.92 0.042
6.92 0.056
6.93 0.07
6.93 0.084
6.94 0.098
6.95 0.111
7.41 0.125
7.43 0.139
7.88 0.153
7.9 0.167
7.91 0.181
7.91 0.195
7.92 0.209
7.92 0.222
7.93 0.236
7.93 0.25
7.94 0.264
7.96 0.278
8.43 0.292
8.89 0.306
8.91 0.32
8.91 0.333
8.92 0.347
8.92 0.361
8.93 0.375
8.93 0.389
8.93 0.403
8.94 0.417
8.94 0.431
8.95 0.444
8.96 0.458
8.98 0.472
9.88 0.486
9.9 0.5
9.91 0.514
9.91 0.528
9.92 0.542
9.92 0.555
9.92 0.569
9.93 0.583
9.93 0.597
9.93 0.611
9.94 0.625
9.95 0.639
9.96 0.653
10.4 0.666
10.9 0.68
10.9 0.694
10.9 0.708
10.9 0.722
10.9 0.736
10.9 0.75
10.9 0.764
10.9 0.777
10.9 0.791
10.9 0.805
11 0.819
11 0.833
11.4 0.847
11.9 0.861
11.9 0.875
11.9 0.888
12.9 0.902
12.9 0.916
13.9 0.93
14.9 0.944
16.4 0.958
16.9 0.972
17.9 0.986
21.9 0.999
32.4 1
};
\addlegendentry{4 TDD periods backward}
\end{axis}

\end{tikzpicture}
    \vspace{-5pt}
    \caption{\gls{rtt} w/ \gls{dl}-aware proactive scheduler at varying lookback windows.}
    \label{fig:delay-aware-RTT}
\end{figure}

Fig.~\ref{fig:delay-aware-RTT} measures \gls{rtt} in a real \gls{5g} network where a \gls{ue} receives periodic pings from the \gls{upf}, and where the scheduler proactively allocates \gls{ul} resources whenever a \gls{dl} transmission occurred within a lookback of $1$--$4$ \gls{tdd} periods. With $2$ periods the median \gls{rtt} drops to ${\sim}9$~ms (i.e., comparable to always-on scheduling) while resource overhead is significantly reduced. A coarse correlation signal (here, recent \gls{dl} transmissions) is therefore already informative enough to bypass the \gls{sr}; the rest of this section generalizes this idea into the full learning framework.

\subsection{System Architecture}

\begin{figure}
    \centering
    \includegraphics[width=1.\linewidth]{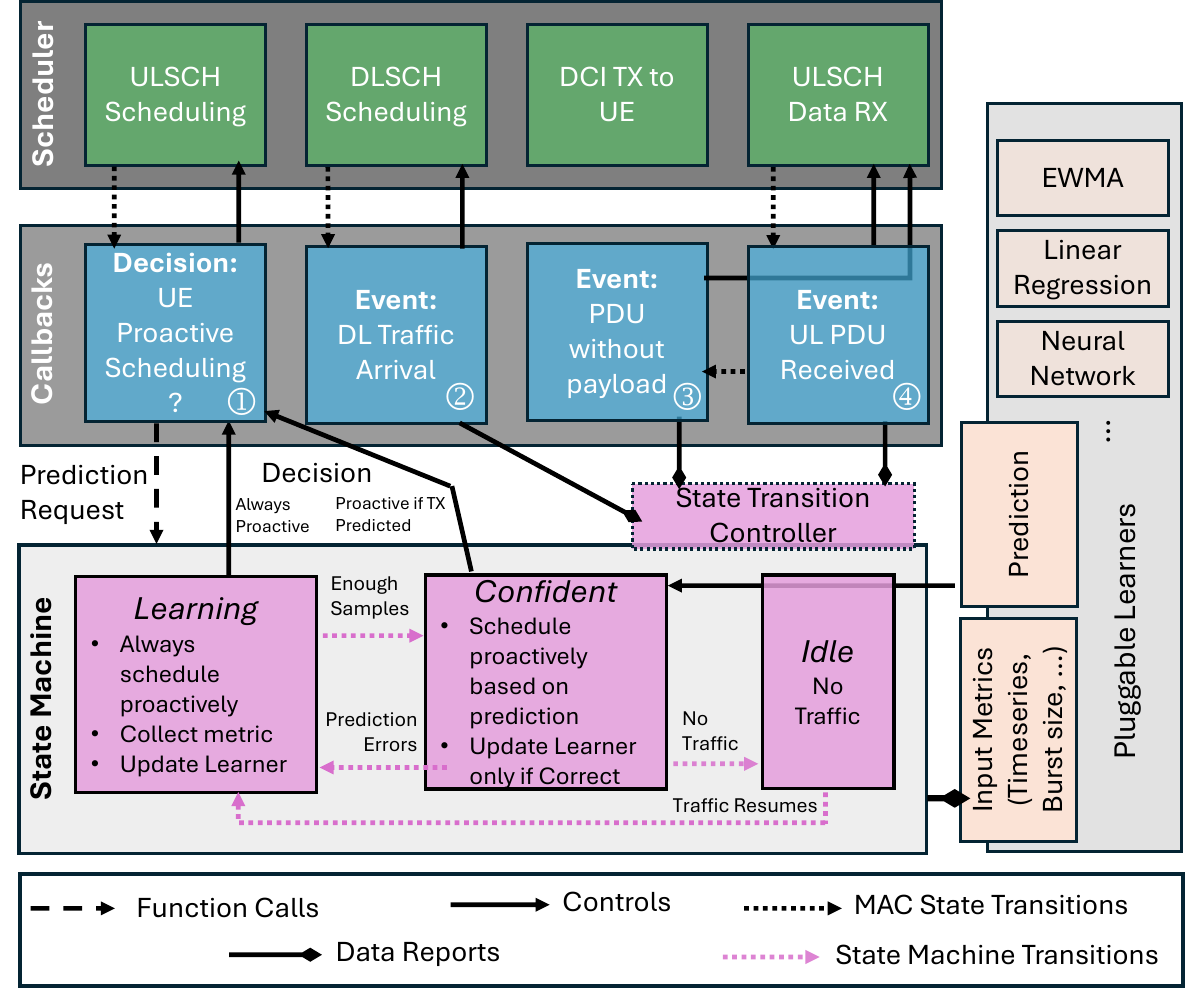}
    \caption{Proposed \framework scheduler framework.}
    \label{fig:framework}
    \vspace{-17pt}
\end{figure}

\framework follows a \gls{cb}-based architecture (Fig.~\ref{fig:framework}): \gls{mac}-layer events from the scheduler are surfaced through \glspl{cb} that feed a state machine, which then modifies the \gls{ulsch} scheduling. \framework instruments three \gls{ulsch} \glspl{cb}. In Fig.~\ref{fig:framework}, \gls{cb}~\cnum{1} is a per-slot proactive query to the state machine, \gls{cb}~\cnum{4} reports \gls{rnti}, \gls{sdu} size, timestamp, and grant type for each received \gls{pdu}, \gls{cb}~\cnum{3} fires when a proactive grant returned empty, and \gls{cb}~\cnum{2} is the only one that happens in the \gls{dlsch}; it reports timestamp and size of each \gls{dl} transmission. Additional \glspl{cb} (e.g., pilot or \gls{harq} reception) can be added without modifying the rest of the framework; the four above suffice for our use cases.

We implement \framework as a set of modifications to the \gls{oai} \gls{gnb} \gls{mac} layer. The underlying allocation policy keeps the \gls{pf} algorithm~\cite{allstar}. \framework's control \gls{cb}~\cnum{1} hooks into the \gls{ul} scheduling loop and inserts the target \gls{ue} into the candidate set of the current slot whenever the state machine triggers a proactive grant. Because the \gls{ue} has not issued an \gls{sr}, no real \gls{bsr} is available, so the \gls{cb} also attaches a virtual non-zero buffer size for the \gls{pf} scheduler to size the grant. The scheduler then allocates \glspl{prb} as it would for any \gls{ue} with pending data, so \gls{mcs} selection, link adaptation, and inter-\gls{ue} fairness behave identically to the baseline. Proactive scheduling changes \emph{who} is eligible in a given slot, not \emph{how} allocation happens, so the framework composes with any underlying scheduling policy. Under \gls{pf} specifically, the two mechanisms reinforce each other: the \gls{pf} metric is the ratio of instantaneous achievable rate to historical throughput, and sparse \gls{urllc} traffic drives the denominator toward zero, so a proactively scheduled \gls{ue} carries an effectively unbounded \gls{pf} score and dominates contention against any non-\gls{urllc} \gls{ue}. Our experiments assume no congestion, which does not always hold in practice; the standard remedy is to group \glspl{ue} and reserve a fraction of the cell's \glspl{prb} for them (i.e., slicing), which we have previously enabled in~\cite{allstar}.

\subsection{State Machine Design}
\label{sec:state-machine}

A state machine instance is associated with each group of \glspl{ue} sharing a common arrival distribution: per-\gls{ue} when no such prior is available, or per-group when \glspl{ue} are known to share statistics (e.g., identical sensors on a fixed reporting schedule). We describe one such instance below; the framework runs them independently, one per group.

At initialization the instance is in the \emph{Learning} state, where it schedules proactively at every slot. This both masks the \gls{sr} delay and collects an unbiased dataset of true \gls{ul} arrival times. In parallel, the learner is updated; it can be any \gls{ml} model (e.g., \gls{ewma}, linear regression, pretrained \gls{nn}) that consumes any subset of scheduler-observable features (past \gls{ul} inter-arrival times, \gls{dl} event sizes and timestamps, \gls{dl}-to-\gls{ul} response times, traffic class) and emits a predicted slot offset for the next \gls{ul} arrival. The feature subset is deployment-specific: periodic reporting may use only past inter-arrival times, while cued edge inference exploits recent \gls{dl} burst sizes. The transition to \emph{Confident} fires once a learner-specific readiness condition is met (e.g., $N$ samples collected or prediction variance below threshold).

In the \emph{Confident} state, scheduling follows the learner's prediction, governed by two operator-tunable knobs: a tolerance window $\mathrm{TW}$ that admits scheduling within $\pm \mathrm{TW}$ \gls{tdd} periods of the prediction, and a slot restriction $N$ that limits proactive grants to the last $N$ slots of each \gls{tdd} period. Smaller $(\mathrm{TW}, N)$ minimize overhead by committing to the predicted slot only; larger values widen the safety window at the cost of additional grants. In the language of~\eqref{eq:multiobj}, the \emph{Learning} state realizes the $\min \max_i \mathbb{E}[L^i_{sched}]$ extreme at the cost of maximal $\sum_i U_i$, while $(\mathrm{TW}, N) = (0,1)$ approaches the opposite extreme (minimum overhead, latency bounded by prediction error). Intermediate values trace the empirical Pareto frontier; the achievable trade-off depends entirely on prediction accuracy.

Prediction quality is monitored online via two counters: false negatives (reactive transmissions, tracked as consecutive misses) and false positives (unused proactive grants, tracked cumulatively). Crossing configurable thresholds (e.g., 3 consecutive misses or a 10\% false-positive rate) triggers a state transition. The framework also supports learners that emit $(\mathrm{TW}, N)$ dynamically, letting the system absorb drift in the arrival distribution by temporarily widening its safety window without fully reverting to \emph{Learning}. A third \emph{Idle} state handles \glspl{ue} with extended inactivity (no \gls{ul} for 100+ slots), pausing prediction until traffic resumes.

Together, the three states let the framework balance latency, jitter, and overhead under unknown and possibly drifting traffic, without \gls{rrc} reconfiguration or a priori knowledge of the arrival process.

\textbf{Scaling Considerations.} The framework's scaling is bounded primarily by \gls{pusch} \gls{prb} consumption per \gls{ul} slot. Each proactive grant consumes a tunable number of \glspl{prb} sized to the expected payload, so the per-slot \gls{prb} budget caps the number of concurrent proactive \glspl{ue}. A $100$~MHz cell carries up to $273$ \glspl{prb} per slot; even the extreme lower bound of $1$ \gls{prb} per grant yields a \gls{tbs} of only a few tens of bytes at typical channel conditions~\cite{ts38214}. The \gls{pdcch}, which carries the grants themselves, is not the binding constraint: a typical CORESET configuration can schedule tens of \glspl{ue} per slot~\cite{mozaffari2021blocking}, and typical \gls{tdd} patterns have more \gls{dl} slots (which carry grants) than \gls{ul} slots (which carry data).

\section{Experimental Setup}
\label{sec:setup}

\begin{figure}
    \centering
    \includegraphics[width=0.95\linewidth]{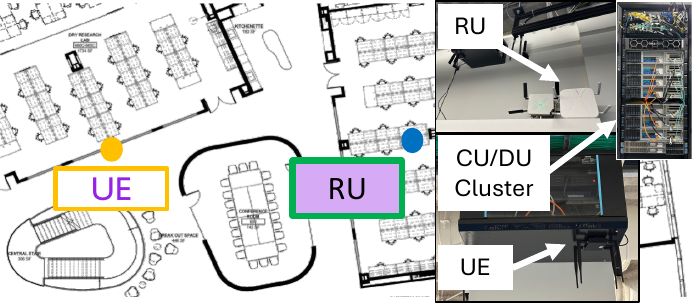}
    \caption{Lab testbed for \framework evaluation.}
    \label{fig:testbed}
    \vspace{-15pt}
\end{figure}

We evaluate \framework on a real \gls{oai}-based \gls{5g} \gls{sa} over-the-air deployment built on the X5G testbed~\cite{villa2024x5g} and the AutoRAN framework~\cite{maxenti2025autoran}. The \gls{gnb} runs on a Dell PowerEdge R760 server connected to a Foxconn RPQN \gls{ru} operating in band n78 with 100~MHz bandwidth and numerology $\mu=1$. A single Sierra Wireless EM9293 \gls{ue} communicates over the air in a lab environment with fair channel conditions (median \gls{cqi}~13). The network operates with a \gls{tdd} pattern of DDDDDDDSUU (10-slot periodicity) typical of commercial deployments. Fig.~\ref{fig:testbed} summarizes the deployment.

\subsection{Evaluation Scenarios}

We design three complementary scenarios that represent key \gls{urllc} use cases where proactive scheduling provides significant benefits.

\textbf{Request-Response Pattern (RTT Measurements).} We generate ICMP echo requests (300-byte payload) from the \gls{upf} to the \gls{ue} at controlled intervals. This scenario represents query-response applications where network elements request data from edge devices, e.g., command-and-control polling of forward assets, \acrshort{fpv} \gls{uav} teleoperation control loops, and analogous civilian workloads such as industrial sensors responding to polling or \gls{v2x} infrastructure querying vehicle status. The predictable \gls{dl}-to-\gls{ul} pattern allows us to evaluate \gls{rtt}-based learning where the scheduler predicts \gls{ul} responses based on prior \gls{dl} transmissions.
In this scenario, the framework is configured to use an \gls{ewma} learner.

\textbf{Cued Edge Inference.} We emulate workloads where a central node queries an edge device, which runs local \gls{ml} inference and returns a variable-size result. This pattern appears in tactical \gls{isr} (cued sensor exploitation, where a forward sensor is tasked to run a focused classifier on a target region), tactical \gls{nlp}/signal intelligence triage (forward transcription or translation of pushed audio), and edge-cloud split \gls{ai} assistants on forward operating bases. The common shape across these use cases is that the inference latency and the size of the resulting \gls{ul} payload both depend on characteristics of the incoming \gls{dl} traffic.
We instantiate this scenario by running an Arctic Embed XS text-embedding model on the \gls{ue}: it receives a variable-length prompt over \gls{dl}, performs inference whose duration depends on input length, and returns the embedding over \gls{ul}, with the server at the \gls{ue} and the client at the \gls{upf} communicating over \gls{udp}. We vary the input by generating random word sequences of different lengths, which translates into different packet sizes at the \gls{mac} layer. \framework is configured to use a Linear Regression on the \gls{dl} burst size for prediction; this is one natural instantiation of the more general principle that the learner can consume any scheduler-observable feature correlated with the \gls{ul} arrival time.

\textbf{Periodic Autonomous Reporting.} We generate periodic \gls{ul}-only traffic at fixed intervals (10-100~ms) with no preceding \gls{dl} trigger, representing workloads such as unattended tactical sensors, periodic status reports from autonomous systems, mission-critical telemetry, and analogous industrial \gls{iot} traffic (heartbeat messages, autonomous sensor reporting). This scenario has no \gls{dl} correlation to exploit, requiring the scheduler to learn pure inter-arrival patterns. It demonstrates the framework's ability to handle traffic that grant-free scheduling targets but without the associated rigidity constraints.
For this experiment, we use mgen to send 97 packets per second. We select this value because it is a prime number, which ensures that the position of packets in the \gls{tdd} period does not repeat itself; it also produces a $10.31$~ms inter-arrival time that falls between the standardized \gls{cg} periodicities ($10$ and $20$~ms), making this scenario representative of periodic workloads that \gls{cg} scheduling cannot serve without phase drift.
In this scenario, we use an \gls{ewma} learner.

For evaluation, we disable the Confident-to-Learning fallback which would normally be triggered by accumulated prediction errors (Sec.~\ref{sec:state-machine}). Otherwise, measured latency and overhead would mix two regimes: the predictor's accuracy in the Confident state, and the always-schedule mode of the Learning state, which by construction sits at the (100\% overhead, minimum delay) corner. Disabling the fallback isolates the first effect, so the reported numbers reflect Confident-state prediction quality alone.
Our baseline for evaluation is the \gls{oai} scheduler which does not perform proactive scheduling.
\section{Performance Evaluation}
\label{sec:results}

For each scenario we evaluate the latency CDFs across the $(\mathrm{TW}, N)$ sweep, together with the overhead (i.e., the fraction of slots in which the \gls{ue} is scheduled) for the request-response case. The sweep traces an empirical latency-overhead trade-off curve, and two observations recur across all three scenarios. First, the curve exhibits a sharp \emph{knee}: a single $(\mathrm{TW}, N)$ configuration achieves near-minimum overhead and near-minimum latency simultaneously, while more aggressive configurations are dominated (i.e., gain nothing on either metric). Second, the location of the knee is an empirical readout of prediction quality. For example, a knee at $(\mathrm{TW}, N) = (0, 1)$ means the learner localizes \gls{ul} arrivals to within a single \gls{tdd} period, so larger safety windows offer no further benefit.

\textbf{Request-Response (RTT).}
We vary $\mathrm{TW}$ between $0$ (proactive scheduling only if the \gls{tdd} period matches the prediction exactly) and $3$ ($\pm 3$ periods around the prediction). We vary $N$ between $1$ and $3$, since we have at most $3$ \gls{ul} slots per \gls{tdd} period in our pattern (two U slots and one mixed S slot).

\begin{figure}
\centering
    \setlength\fwidth{0.9\linewidth}
    \setlength\fheight{2cm}
    \input{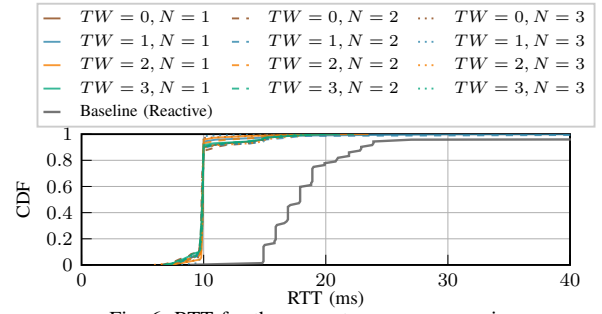}
    \vspace{-.3cm}
    \caption{\gls{rtt} for the request-response scenario.}
    \label{fig:latency-rtt}
    \vspace{-13pt}
\end{figure}

Fig.~\ref{fig:latency-rtt} presents the latency results. For every $(\mathrm{TW}, N)$, the latency distribution improves substantially over the baseline: the median drops from $\sim 20$~ms with a long tail to $10$~ms, with a 90th percentile also at $10$~ms and no long tail.

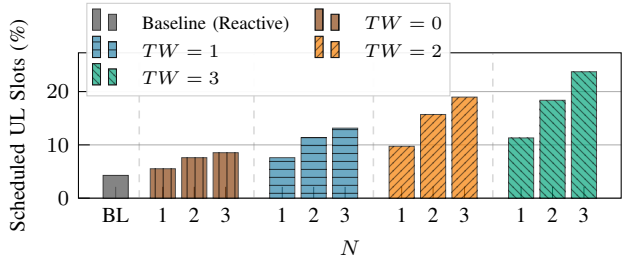
\begin{figure}[!htb]
\centering
    \setlength\fwidth{0.9\linewidth}
    \setlength\fheight{2cm}
\begin{tikzpicture}
\usetikzlibrary{patterns}
\definecolor{darkcyan6167125}{RGB}{6,167,125}
\definecolor{darkgrey176}{RGB}{176,176,176}
\definecolor{darkorange2471270}{RGB}{247,127,0}
\definecolor{dimgrey102}{RGB}{102,102,102}
\definecolor{grey}{RGB}{128,128,128}
\definecolor{lightgrey204}{RGB}{204,204,204}
\definecolor{saddlebrown}{RGB}{139,69,19}
\definecolor{steelblue46134171}{RGB}{46,134,171}

\begin{axis}[
width=\linewidth,
height=0.4\linewidth,
legend cell align={left},
legend style={
  fill opacity=0.95,
  draw opacity=1,
  text opacity=1,
  draw=lightgrey204,
  at={(0.35,0.7)},
  anchor=south,
  legend columns=2,  
  column sep=0.3cm,  
  font=\scriptsize,       
  inner sep=2pt,     
  nodes={inner sep=1pt}  
},
tick align=inside,
tick pos=left,
x grid style={darkgrey176},
xmin=-1.185, xmax=16.085,
xtick style={color=black},
xtick={0,1.5,2.5,3.5,5.3,6.3,7.3,9.1,10.1,11.1,12.9,13.9,14.9},
xticklabels={BL
 ,1,2,3,1,2,3,1,2,3,1,2,3},
y grid style={darkgrey176},
ylabel={Scheduled UL Slots (\%)},
ymajorgrids,
ymin=0, ymax=27.2791046724093,
ytick style={color=black},
label style={font=\footnotesize},
ticklabel style={font=\footnotesize},
xlabel={$N$}
]
\draw[draw=black,fill=dimgrey102,opacity=0.8,line width=0.28pt] (axis cs:-0.4,0) rectangle (axis cs:0.4,4.27142857142857);
\addlegendimage{ybar,ybar legend,draw=black,fill=dimgrey102,opacity=0.8,line width=0.28pt}
\addlegendentry{Baseline (Reactive)}

\draw[draw=black,fill=saddlebrown,opacity=0.7,line width=0.28pt,postaction={pattern=vertical lines, fill opacity=0.7}] (axis cs:1.1,0) rectangle (axis cs:1.9,5.52090330422544);
\addlegendimage{ybar,ybar legend,draw=black,fill=saddlebrown,opacity=0.7,line width=0.28pt,postaction={pattern=vertical lines, fill opacity=0.7}}
\addlegendentry{$TW = 0$}

\draw[draw=black,fill=saddlebrown,opacity=0.7,line width=0.28pt,postaction={pattern=vertical lines, fill opacity=0.7}] (axis cs:2.1,0) rectangle (axis cs:2.9,7.59710840162569);
\draw[draw=black,fill=saddlebrown,opacity=0.7,line width=0.28pt,postaction={pattern=vertical lines, fill opacity=0.7}] (axis cs:3.1,0) rectangle (axis cs:3.9,8.53325336532054);
\draw[draw=black,fill=steelblue46134171,opacity=0.7,line width=0.28pt,postaction={pattern=horizontal lines, fill opacity=0.7}] (axis cs:4.9,0) rectangle (axis cs:5.7,7.59897205686217);
\addlegendimage{ybar,ybar legend,draw=black,fill=steelblue46134171,opacity=0.7,line width=0.28pt,postaction={pattern=horizontal lines, fill opacity=0.7}}
\addlegendentry{$TW = 1$}

\draw[draw=black,fill=steelblue46134171,opacity=0.7,line width=0.28pt,postaction={pattern=horizontal lines, fill opacity=0.7}] (axis cs:5.9,0) rectangle (axis cs:6.7,11.3582114641116);
\draw[draw=black,fill=steelblue46134171,opacity=0.7,line width=0.28pt,postaction={pattern=horizontal lines, fill opacity=0.7}] (axis cs:6.9,0) rectangle (axis cs:7.7,13.1414141414141);
\draw[draw=black,fill=darkorange2471270,opacity=0.7,line width=0.28pt,postaction={pattern=north east lines, fill opacity=0.7}] (axis cs:8.7,0) rectangle (axis cs:9.5,9.73569023569024);
\addlegendimage{ybar,ybar legend,draw=black,fill=darkorange2471270,opacity=0.7,line width=0.28pt,postaction={pattern=north east lines, fill opacity=0.7}}
\addlegendentry{$TW = 2$}

\draw[draw=black,fill=darkorange2471270,opacity=0.7,line width=0.28pt,postaction={pattern=north east lines, fill opacity=0.7}] (axis cs:9.7,0) rectangle (axis cs:10.5,15.7040731504572);
\draw[draw=black,fill=darkorange2471270,opacity=0.7,line width=0.28pt,postaction={pattern=north east lines, fill opacity=0.7}] (axis cs:10.7,0) rectangle (axis cs:11.5,18.9655172413793);
\draw[draw=black,fill=darkcyan6167125,opacity=0.7,line width=0.28pt,postaction={pattern=north west lines, fill opacity=0.7}] (axis cs:12.5,0) rectangle (axis cs:13.3,11.3036303630363);
\addlegendimage{ybar,ybar legend,draw=black,fill=darkcyan6167125,opacity=0.7,line width=0.28pt,postaction={pattern=north west lines, fill opacity=0.7}}
\addlegendentry{$TW = 3$}

\draw[draw=black,fill=darkcyan6167125,opacity=0.7,line width=0.28pt,postaction={pattern=north west lines, fill opacity=0.7}] (axis cs:13.5,0) rectangle (axis cs:14.3,18.3694374475231);
\draw[draw=black,fill=darkcyan6167125,opacity=0.7,line width=0.28pt,postaction={pattern=north west lines, fill opacity=0.7}] (axis cs:14.5,0) rectangle (axis cs:15.3,23.7209605847037);
\addplot [semithick, grey, opacity=0.3, dashed, forget plot]
table {%
0.75 0
0.75 27.2791046724093
};
\addplot [semithick, grey, opacity=0.3, dashed, forget plot]
table {%
4.4 0
4.4 27.2791046724093
};
\addplot [semithick, grey, opacity=0.3, dashed, forget plot]
table {%
8.2 0
8.2 27.2791046724093
};
\addplot [semithick, grey, opacity=0.3, dashed, forget plot]
table {%
12 0
12 27.2791046724093
};
\end{axis}

\end{tikzpicture}
        \vspace{-5pt}
    \caption{Resource overhead for the request-response scenario; bottom numbers indicate the slot restriction $N$.}
    \label{fig:overhead}
    \vspace*{-15pt}
\end{figure}

As depicted in Fig. \ref{fig:overhead}, the overhead (measured as the proportion of \gls{ul} slots with some resources scheduled, either proactively or reactively) scales linearly with $\mathrm{TW}$ and $N$. Hence, the optimal configuration for this scenario is $\mathrm{TW}=0$, $N=1$, where the overhead is 7\%. On the other hand, the baseline has the lowest overhead with 4\% of slots scheduled.
While we present detailed overhead analysis only for the request-response scenario, the overhead scales similarly with $\mathrm{TW}$ and $N$ across all scenarios, with absolute values proportional to the traffic rate; we omit the overhead plots for the other scenarios for brevity.

\begin{figure}
\centering
    \setlength\fwidth{0.9\linewidth}
    \setlength\fheight{2cm}
    \input{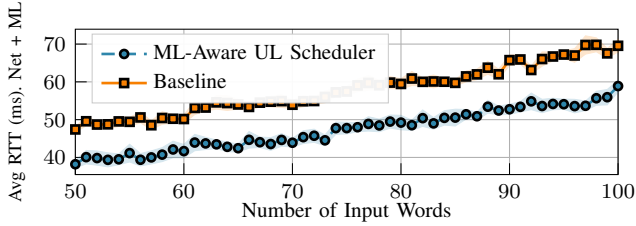}
    \vspace{-8pt}
    \caption{E2E latency for the cued edge inference scenario.}
    \label{fig:ml_inference}
    \vspace{-12pt}
\end{figure}

\textbf{Cued Edge Inference.}
We use $\mathrm{TW}=2$ and $N=1$. As shown in Fig.~\ref{fig:ml_inference}, the scheduler predicts the \gls{ul} arrival time from the \gls{dl} burst size, giving an end-to-end delay reduction of $\sim 10$~ms over the baseline across all input sizes, in line with the network latency saved in the request-response scenario.

\begin{figure}[!htb]
\vspace{-8pt}
\centering
    \setlength\fwidth{0.9\linewidth}
    \setlength\fheight{2cm}
    \input{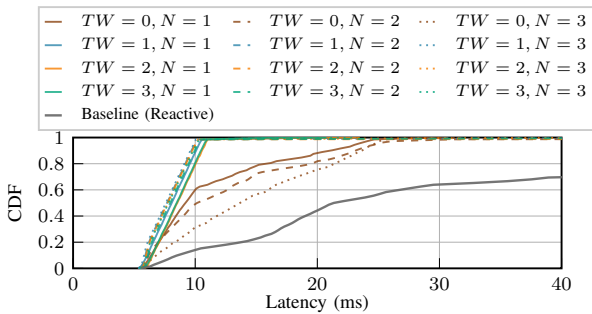}
    \vspace{-5pt}
    \caption{One way latency for the uplink autonomous reporting.}
    \label{fig:oneway}
    \vspace{-15pt}
\end{figure}

\textbf{Periodic Autonomous Reporting.}
Fig.~\ref{fig:oneway} presents the third scenario. As before, we vary $\mathrm{TW}$ and $N$ between $0$ and $3$. $\mathrm{TW}=0$ already improves the CDF over the baseline, but the best results appear from $\mathrm{TW}=1$, with a median one-way latency of $7$~ms and a 99\textsuperscript{th} percentile of $10$~ms. The shift in optimal $\mathrm{TW}$ reveals a difference in detected traffic pattern for this use case. The prime $97$-packet/s rate produces a $10.31$~ms inter-arrival that is not a multiple of the \gls{tdd} period, so successive packets walk across slot boundaries within the pattern. A predictor with $\mathrm{TW}=0$ commits to a single \gls{tdd} period and misses the arrival as drift accumulates; $\mathrm{TW}=1$ absorbs the drift at low overhead. \gls{cg} cannot apply the same optimization: its fixed periodicities ($10$ and $20$~ms) lock the grant to a single phase, so the same drift that $\mathrm{TW}=1$ tolerates would steadily desynchronize a \gls{cg} schedule from the actual arrivals. The difference of up to a tenth of a millisecond between $N=1, 2, 3$ comes from the high regularity of the periodic generation, where scheduling a few slots earlier has a small but noticeable impact on \gls{ul} latency. The optimal operating point on the Pareto frontier is hence traffic-dependent; as real traffic is dynamic, \framework supports learners that emit $(\mathrm{TW}, N)$ from observed statistics.

\section{Conclusions}
\label{sec:conclusion}

This paper presented \framework, a learning-based \gls{mac} scheduling framework that addresses the latency--efficiency trade-off in \gls{5g} \gls{urllc} systems. By embedding \gls{ml} models in the \gls{ul} scheduler and supervising them with an adaptive state machine, we eliminate the \gls{sr}-induced delay and keep radio overhead close to the \gls{sr}-based baseline. The integration with \gls{oai} leaves the underlying \gls{pf} allocation policy unchanged, so the framework only modifies which \glspl{ue} are eligible in a given slot and composes with any scheduling policy.

Across three \gls{urllc} scenarios on a \gls{5g} testbed, \framework reduces median \gls{rtt} by $\approx 50\%$ at $7$--$10\%$ radio overhead, and reduces median one-way latency from $\sim 13$~ms to $7$--$10$~ms. The empirical $(\mathrm{TW}, N)$ sweeps trace a Pareto frontier of the latency--overhead trade-off with a sharp knee, showing that the learner localizes \gls{ul} arrivals accurately enough that larger safety windows offer no further benefit. \framework therefore operates at the best achievable point on this trade-off.

Future work includes multi-\gls{ue} experiments under congestion, online adaptation of $(\mathrm{TW}, N)$, and more expressive learners for traffic with multiple superimposed periodicities. On the channel side, each grant commits an \gls{mcs} up to $\mathrm{TW}$ periods ahead, so we plan to predict channel evolution from sparse feedback (\gls{srs} / \gls{csi}). Finally, the implementation-level optimizations of~\cite{gong2025towardsurllc} are orthogonal to \framework: they eliminate the \gls{sr} via static per-slot reservation (the always-on extreme of our Pareto frontier), whereas \framework eliminates it adaptively at ${\sim}10\%$ overhead. Combining the two would compound the lower-level latency reductions with our prediction-driven overhead savings.
\vspace{-3pt}
\bibliographystyle{IEEEtran}
\bibliography{refs}

\end{document}